\documentclass[%
 aip,
 jmp,%
 amsmath,
 amssymb,
 twocolumn,
 floatfix,
]{revtex4-2}
\usepackage{graphicx}% Include figure files
\usepackage{dcolumn}% Align table columns on decimal point
\usepackage{bm}% bold math
\usepackage{todonotes} % TODO notes for draft
\usepackage[inkscapelatex=false]{svg} % For vector graphic figures
\usepackage[colorlinks=true, allcolors=blue]{hyperref} % Pretty links to references and citations
\usepackage[normalem]{ulem}   % To get sout command for strikeout
\usepackage{array}

\makeatletter
\renewcommand*{\eqref}[1]{%
  \hyperref[{#1}]{\textup{\tagform@{\ref*{#1}}}}%
}

\makeatother
\begin{document}
\title{Exploring Direct Detection of Massive Particles Using Wave Propagation from Gravitational Coupling with a Wire Under Tension}
\author{Thomas Belvin}
\affiliation{Physics Department, University of Maryland College Park}
\author{Peter Shawhan}
\affiliation{Physics Department, University of Maryland College Park}

\date{\today}

\begin{abstract}
We investigate the feasibility of detecting galactic orbit dark matter passing through Earth by measuring its gravitational coupling with a wire under tension. 
We do so by exploring the transverse and longitudinal waves induced on the wire to detect a massive particle passing within $\sim 1$ m of the wire. 
The particle's $r^{-2}$ interaction with the wire provides an initial momentum which develops into a propagating wave carrying a distinctive time dependent displacement. 
Most interestingly, we find that both transverse and longitudinal waves develop with unique profiles, allowing for a full, three dimensional reconstruction of the particle's trajectory and its mass over velocity ratio. 
We find that, at interaction distances of 0.1 to 100 mm with a 90 micron diameter copper beryllium wire, Planck scale dark matter with mass $\sim 10^{19}$ GeV/$c^2$ would create immeasurable displacements on the scale of $10^{-24}$ to $10^{-26}$ m. 
In order to create displacements detectable by modern, commercially available, displacement sensors on the nanometer scale we require dark matter with a particle mass greater than $4 \times 10^7$ kg ($\sim 2 \times 10^{32}$ GeV/$c^2$). 
This is outside the upper limit of the Planck scale by 13 orders of magnitude and would also have such a low particle flux that a detection event would be implausible.
Finally, we perform a similar analysis for a charged wire and an elementary charged particle with their electrostatic interaction, finding that a sufficiently slow charged particle would produce a transverse displacement comparable to the sensitivity of currently available sensors.
\end{abstract}


\maketitle
\section{Introduction}
% What is dark matter
Observations such as the galactic orbital velocity curve lead us to believe there exists dark matter which interacts most prominently through gravitational coupling, with at most a very weak interaction with other forces.\cite{Billard_2022, Schumann_2019, Liu2017, Drees_2012, JUNGMAN1996195} 
Weakly interacting massive particles, or WIMPS, have been the subject of much research and experiments aimed at indirect or direct detection of their interactions with traditional particles and forces. \cite{Billard_2022, Schumann_2019, Liu2017, arcadi2024waningwimpendgame}
The most general, but also most challenging, detection prospect for dark matter is to rely solely on the gravitational interaction. 
For example, the Windchime Project has outlined an ambitious plan to use a matrix of point-like displacement sensors to detect and reconstruct interaction with traveling dark matter.\cite{Carney2020,WindchimeWebsite} 
This paper investigates the same question instead using a wire as the sensor for the gravitational interaction with a dark matter particle. 
We additionally perform similar analyses of the gravitational interaction of thermal neutrons and the electrostatic interaction of charged particles with a charged wire.

% Why this work is different and worth publishing
When encountered by a passing massive particle, the wire gains momentum through the gravitational interaction between itself and the passing particle which will develop into traveling pulses of displacement that can then be detected by displacement sensors. 
We chose to investigate this detection method because of the mechanical simplicity and potential for cheap production costs compared to the sensitive area of the detector. 
The sensitive area of the interaction benefits because the wire acts as a continuous series of point-like pick-ups for the gravitational impulse of the passing particle, increasing the cross sectional area in which a particle can be detected compared to a point-like sensor.

% Particle properties
Any device designed to detect dark matter particles will respond proportionally to the ratio of the incoming particle's mass to its speed, $M/v$ (See Sec.\ \ref{sec:impulseCalc}).
Therefore, it is necessary to define constraints on a possible velocity for a particle traveling within our galaxy. 
From observations of rotational velocities within the Milky Way, we know that the galactic orbital velocity near our sun is 230 km/s,\cite{Mroz_2018} and that dark matter is likely responsible for maintaining that velocity out to large radii.\cite{Billard_2022, Schumann_2019, Drees_2012, JUNGMAN1996195} 
We can assume that a dark matter particle moving through the Earth's frame will have a random orbital plane relative to the Milky Way and thus the expected speed relative to Earth will average to 230 km/s.

Additionally, we will want to look for dark matter particles with the greatest mass as they will provide the most displacement. The Planck mass, $m_{Pl} \equiv \sqrt{\hbar c / G}$, is a natural upper limit for elementary dark matter particles. It is numerically equal to about $20\text{ }\mu$g, or $\sim 10^{19}$ GeV/$c^2$ in energy-based units.\cite{Billard_2022, Adhikari2022, Aalbers2024, Aprile2023}

% Observations of the local galactic rotation curve 
While the distribution of dark matter in and around galaxies remains a subject of active research, its local energy density has been estimated to be up to $\rho_{DM}$ = 0.5 GeV/cm$^{3}$.\cite{FabrizioNesti_2013, Schumann_2019, famaey2015darkmattermilkyway, deSalas_2021}
Using this value, and the galactic orbital speed, we can find the expected frequency of events for dark matter particles with specific mass (Sec.\ \ref{sec:GODM}).

\section{Impulse On a Wire From a Passing Particle}\label{sec:impulse}
% What are the interaction requirements
The key to the wire acting as a medium for detection is that its interaction with the dark matter particle provides each element of the wire an initial velocity.

% Calculations for impulse equations
In Newton's theory of universal gravitation, the force felt by a differential segment of wire, with mass $\delta m$, from a massive particle, of mass $M$ whose position relative to the wire segment is given by ${\bm r}$, can be described using the inverse square law times $GM \delta m$,
\begin{equation}
    \bm{F} = GM\delta m \frac{\bm{\hat r}}{r^2} = GM\delta m \frac{\bm{r}}{r^3},
    \label{eqn:general_force}
\end{equation}
where $G$ is the gravitational constant. 
We approximate the wire as a line of mass density assuming that the particle is always at a distance greater than the wire's radius.

For our purposes, the particle speed, $v$, must be fast enough that we can make two more approximations.
First, we assume that the momentum exchanged with the wire is small enough that the particle's trajectory (or track) can be approximated as a straight line.
Second, we assume that the wire will undergo negligible displacement during the interaction, {\it i.e.}, the interaction can be modeled as an impulse. In order for this to be the case, we require the particle speed, $v$, to be much greater than the wave propagation speed of the wire, $w$.
These assumptions simplify our calculation of the impulse given to each differential segment of the wire, which is shown in Section \ref{sec:impulseCalc}.

\subsection{Coordinate System and Particle Trajectory}\label{sec:coordSys}
\begin{figure}[ht!]
    \includegraphics[width=\linewidth]{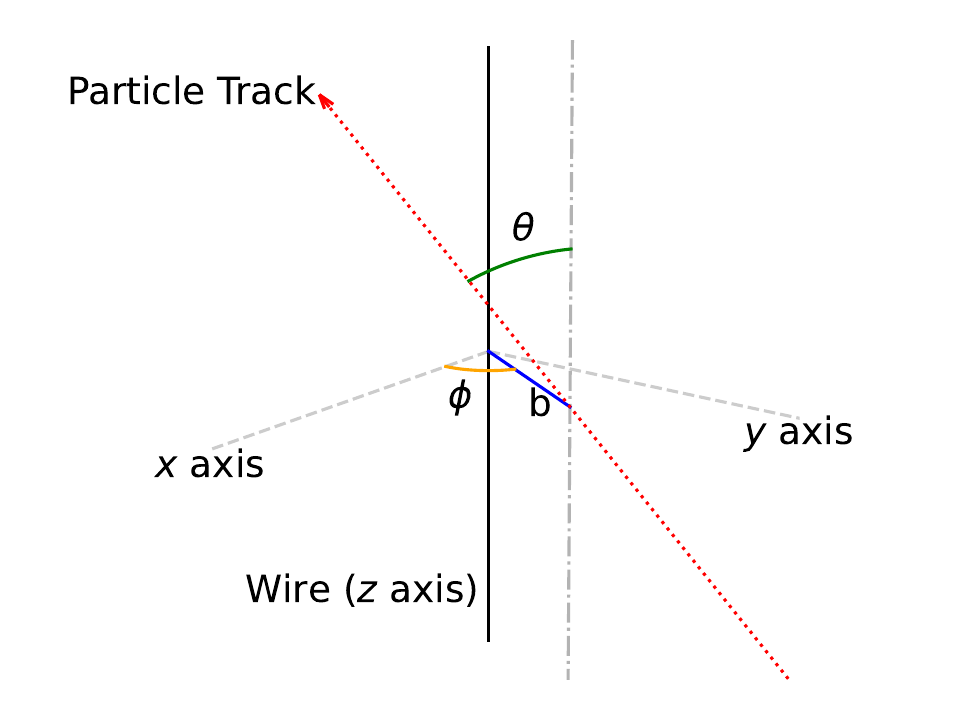}
    \caption{Coordinate system of wire with an over-plotted particle track in dotted red. 
    For this figure $\phi = \pi / 3$ and $\theta = \pi/6$.}
    \label{fig:coordSys}
\end{figure}

To calculate the impulse given to the wire, we use a coordinate system in which the wire lies on the $\bm{\hat z}$ axis and the point of closest approach between the wire and the particle track is centered at $z = 0$ with distance $b$, referred to as the impact parameter from here on. 
We define the angle $\theta$ between the wire ($\bm{\hat z}$ axis) and the particle track, and the angle $\phi$ between the  $\bm{\hat x}$ axis and the impact parameter. 
An example particle track with those coordinates marked is shown in Fig.\ \ref{fig:coordSys}. 

Each differential segment of the wire, labeled by its position $z$, will feel an attractive force along the vector, $\bm r$, due to the passing particle.
Defining $t=0$ to be the moment of closest approach, the vector $\bm{r}$ from a point on the wire, $z$, to the particle's position along its track at any given time, $t$, is given by:
\begin{align}
    \bm{r}=\text{ }
        \nonumber &(b \cos (\phi )+v t \sin (\theta ) \sin (\phi )) \bm{\hat x} \\
        \nonumber + &(b \sin (\phi )-v t \sin (\theta ) \cos (\phi )) \bm{\hat y} \\
                  + &(v t \cos (\theta )-z) \bm{\hat z}.
    \label{eqn:rphi}
\end{align}

This can be greatly simplified by relating it to a track with $\phi = 0$ using a rotation matrix about the z axis, $\bm{R_z}(\phi)$,
\begin{equation}
\bm{R_z}(\phi) = \left(\begin{matrix}
    \cos{\phi} & -\sin{\phi} & 0\\
    \sin{\phi} & \cos{\phi} & 0\\
    0 & 0 & 1
    \end{matrix}\right).
    \label{eqn:RotMat}
\end{equation}
With $\phi = 0$ this leaves the vector $\bm{r}$ as
\begin{equation}
    \bm{r}= b \bm{\hat x} - v t \sin (\theta ) \bm{\hat y} + (v t \cos (\theta )-z) \bm{\hat z}.
    \label{eqn:r}
\end{equation}

\subsection{Impulse Calculations} \label{sec:impulseCalc}
We now calculate the impulse, $\bm{I}$, given to each element of the wire from our inciting particle. 
This is parameterized by $z$, the distance along the wire-axis from the impact parameter. 

Integrating the force from Eq.\ \eqref{eqn:general_force} over all time, we get
\begin{align}
    \nonumber\bm{I}= \text{ } 
    & \frac{2 G M \delta m}{v}\frac{b}{b^2+z^2 \sin ^2(\theta )} \bm{\hat x}\\
    \nonumber            
    - & \frac{2 G M \delta m}{v} \frac{z \sin (\theta ) \cos (\theta )}{b^2+z^2 \sin ^2(\theta )} \bm{\hat y}\\
    - & \frac{2 G M \delta m}{v} \frac{z \sin ^2(\theta )}{b^2+z^2 \sin ^2(\theta )} \bm{\hat z}.
    \label{eqn:impulse}
\end{align}

\begin{figure*}[ht!]
    \centering
    \includegraphics[width=\textwidth]{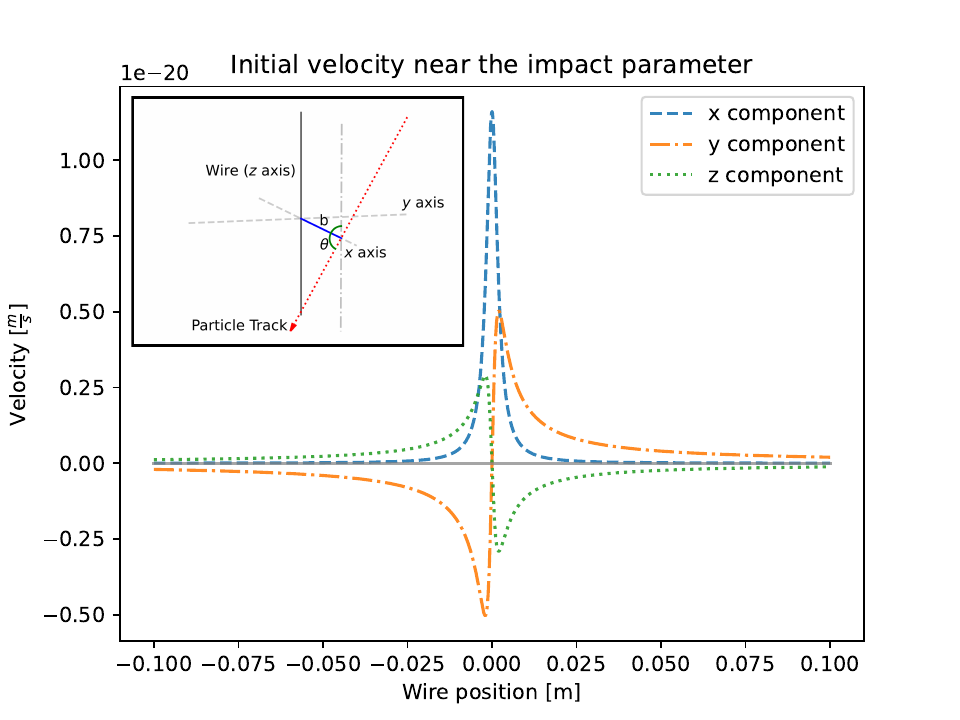}
    \caption{The $x$, $y$, and $z$ components of the initial velocity along the wire within 100 mm of the impact parameter as a function of  $z$ with $b = 1$ mm and $\theta = 5 \pi / 6$. 
    The inset figure depicts the particle track relative to the coordinate system previously defined. The vertical scales in this figure and subsequent figures have been calculated assuming a Planck mass scale dark matter particle in a galactic orbit as described in Sec. \ref{sec:detectableParts}.
    }% Example impulse of a GODM particle passing within 0.1 mm of the center of the wire
    \label{fig:impulseEx1}
\end{figure*}

Figure \ref{fig:impulseEx1} shows the initial velocity on the wire due to the impulse produced by a particle track with $\theta = 5\pi/6$ on each differential segment of the wire labeled by its position $z$. 
The initial velocity is simply the impulse across the wire divided by the mass of the differential wire segments, $\delta m$ and therefore the shape of the velocity and impulse are the same for a wire of uniform mass density. 

We can see that the x-component of the impulse, $I_x$, follows a Lorentzian, as every parameter except $z$ is a constant. 
Meanwhile, the y-component, $I_y$, and z-component, $I_z$, share similar shapes to each other as they only differ between $\sin\theta$ and $\cos\theta$ in the numerator. 
These differing factors give $I_y$ and $I_z$ different maximums in addition to opposite signs when $\sin\theta$ and $\cos\theta$ have opposite signs. 

We note the following features about the impulse: 
\begin{itemize}
    \item First, lower particle speeds result in greater impulse because $v$ is in the denominator.
    \item Second, if $\phi \neq 0$ then $I_x$ and $I_y$ will be superpositions of $I_x$ and $I_y$ for $\phi = 0$ as determined by the rotation matrix (Eq.\ \eqref{eqn:RotMat}). 
    \item Third, if $\theta = 0$ or $\pi$ then the particle track is parallel to the wire and $I_z = 0$ because the net impulse in the z component delivered to any given point by the particle is zero.
    \item Fourth, if $\phi = 0$ and $\theta = \pi / 2$ then the particle track is parallel to the $y$ axis and $I_y = 0$ because $F_y$ is an odd function of $t$ as seen by the behavior of $r_y$ in Eq.\ \eqref{eqn:r}. 
    \item Fifth, if $\phi = \pi / 2$ and $\theta = \pi / 2$ then the particle track is parallel to the $x$ axis and $I_x = 0$  by the same logic as $I_y$ above.
    \item Finally, $I_x$ approaches $0$ as a function of $z$ much faster compared to $I_y$ and $I_z$ due to lacking a factor of $z$ in the numerator.
\end{itemize}

The impulse provides an initial momentum to each element of the wire. 
As we will see, this momentum will create displacement pulses which travel along the wire as transverse and longitudinal waves. 
These waves will fit an expected family of solutions that can be found by solving the wave equation.

\section{Wave Solutions for a Wire with Constant Tension}\label{sec:waveSolutions}
% Wave calculations and displacement
We determine the waves which will be created by the impulse derived in Sec.\ \ref{sec:impulseCalc} by finding the solutions whose time derivatives match the initial velocity from the impulse. 
The wire's displacement is described by a 3D vector that depends on $z$ and $t$. 
As a wave in a linear, non-dispersive medium, the general displacement $\bm{\psi}(z,t)$ can be separated into left- and right-traveling components with the same wave speed $w$.
Since the entire wire has no initial displacement, the left- and right-traveling wave components must have opposite amplitude and thus cancel each other's displacements at $t = 0$:
\begin{equation}
    \bm{\psi_L}(z,0) = -\bm{\psi_R}(z,0).
\end{equation}
Additionally, because we have two pulses traveling in opposite directions, they will each inherit half of the initial momentum, meaning the initial velocities are given by 
\begin{equation}
        \bm{\dot\psi_L}(z, 0) = \bm{\dot\psi_R}(z, 0) = \frac{\bm{I}}{2\, \delta m} \,.
        \label{eqn:wavesHalfImp}
    \end{equation}

The speed of the waves along the wire is determined by the properties of the wire: given a wire of uniform mass density, $\rho$, constant, uniform tension, $T$, and Young's modulus $E$, the transverse wave speed will be
\begin{equation}
    w_t = \sqrt{\frac{T}{\mu}} \,,
    \label{eqn:transWaveSpeed}
\end{equation}
where $\mu$ is the linear mass density of the wire.
The longitudinal wave speed will be
\begin{equation}
    w_l = \sqrt{\frac{E}{\rho}} \,.
\label{eqn:longWaveSpeed}
\end{equation}
For typical metal wires, $w_l>w_t$.

Replacing $z$ with $z \pm w t$, where $w$ is the wave speed, and integrating Eq.\ \eqref{eqn:wavesHalfImp} with respect to time, we get the following equations for the left and right displacement wave solutions:
    \begin{widetext}
        \begin{align} 
        \nonumber \bm{\psi_L}(z + w t) =
            &+ \frac{GM}{v w} \csc (\theta ) \tan ^{-1}\left(\frac{\sin (\theta ) (z + wt)}{b}\right) \bm{\hat x}\\ \nonumber
            &- \frac{GM}{2 v w} \cot (\theta ) \log \left(b^2+\sin ^2(\theta ) (z + wt)^2\right) \bm{\hat y} \\ 
            &- \frac{GM}{2 v w} \log \left(b^2+\sin ^2(\theta ) (z + wt)^2\right) \bm{\hat z} 
            \label{eqn:dispLeft}
    \end{align}
and
    \begin{align} 
        \nonumber \bm{\psi_R}(z - w t) = 
            &- \frac{GM}{v w} \csc (\theta ) \tan ^{-1}\left(\frac{\sin (\theta ) (z-wt)}{b}\right)\bm{\hat x}\\ \nonumber
            &+ \frac{GM}{2 v w} \cot (\theta ) \log \left(b^2+\sin ^2(\theta ) (z-wt)^2\right) \bm{\hat{y}}\\
            &+ \frac{GM}{2 v w} \log \left(b^2+\sin ^2(\theta ) (z-wt)^2\right) \bm{\hat{z}}. 
            \label{eqn:dispRight}
    \end{align}
     \end{widetext}
Note that every component has a factor of $G M/ (v w)$ and thus lower wave speeds give larger displacements. 
This is important because the magnitude of displacement will ultimately determine what interactions can be detected.

\begin{figure}[t!]
    \centering
    \includegraphics[width=\linewidth]{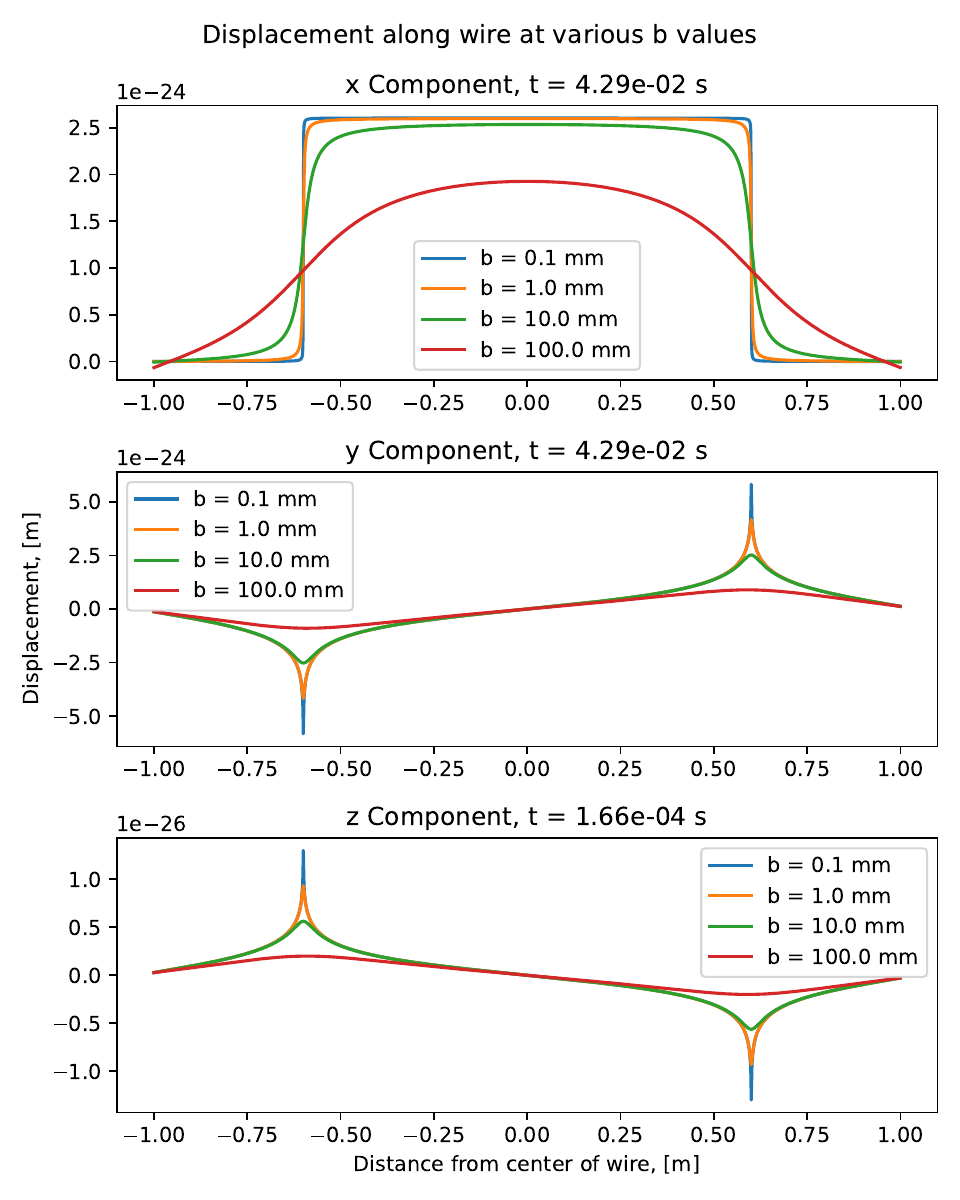}
    \caption{Displacement on wire for several values of b given a galactic orbit Planck scale dark matter particle traveling the same path as shown in Fig.\ \ref{fig:impulseEx1}.}
    \label{fig:diffbsWire}
\end{figure}

%Explanation of diffbsWire showing wire evolution with time to give reader idea 
Figure \ref{fig:diffbsWire} depicts the displacement along the wire at a moment when the pulse is about half way to the end of the wire for four different values of the impact parameter, $b$. 
We can see how the displacement pulses move away from the origin and the overall shape of the displacement they give to the wire. 
Note how the $x$ component remains at its maximum displacement after the pulse has passed while the $y$ and $z$ components have traveling peaks.

\begin{figure}[t!]
    \centering
    \includegraphics[width=\linewidth]{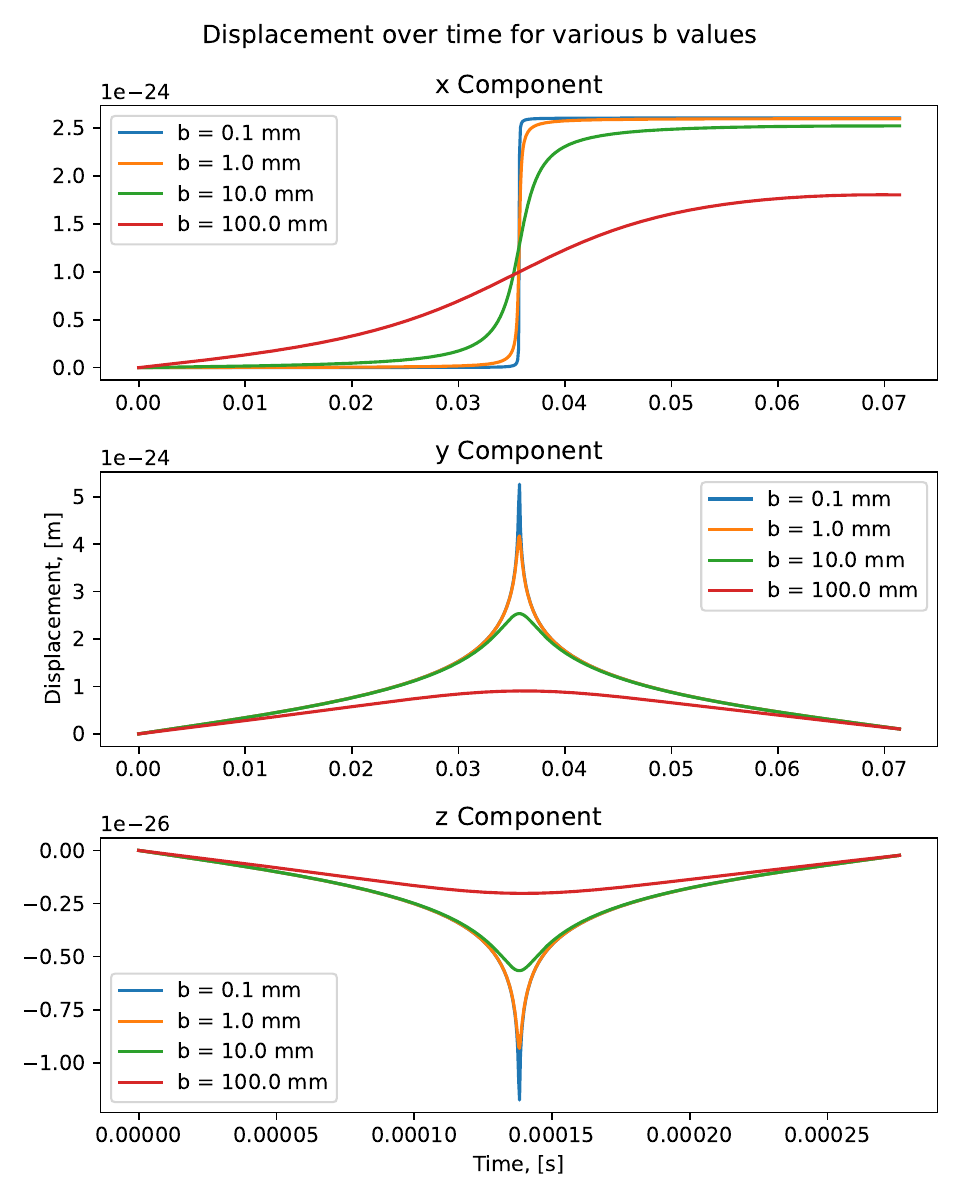}
    \caption{Time evolution of a single point on the wire for various values of b given a galactic orbit Planck scale dark matter traveling the same path as shown in Fig.\ \ref{fig:impulseEx1}.}
    \label{fig:diffbsTime}
\end{figure}

%Explanation of diffbsTime to show what measured data might look like.
Figure \ref{fig:diffbsTime} shows the same displacement pulses but at a single point on the wire over the time it takes the pulse to completely pass that point. 
This is important to look at because if our displacement sensors are placed at a single point along the wire we would expect the data to fit these curves. 

While maximum displacement changes with $b$, the shape is also affected, meaning we can fit the track properties ($\theta$, $\phi$, and $b$) from the pulse shape and use the amplitude to determine the particle's $M/v$ ratio.

\section{Feasibility of Detecting Different Particles}\label{sec:detectableParts}
\subsection{Wire Choice}
\begin{figure}[ht!]
    \centering
    \includegraphics[width=\linewidth]{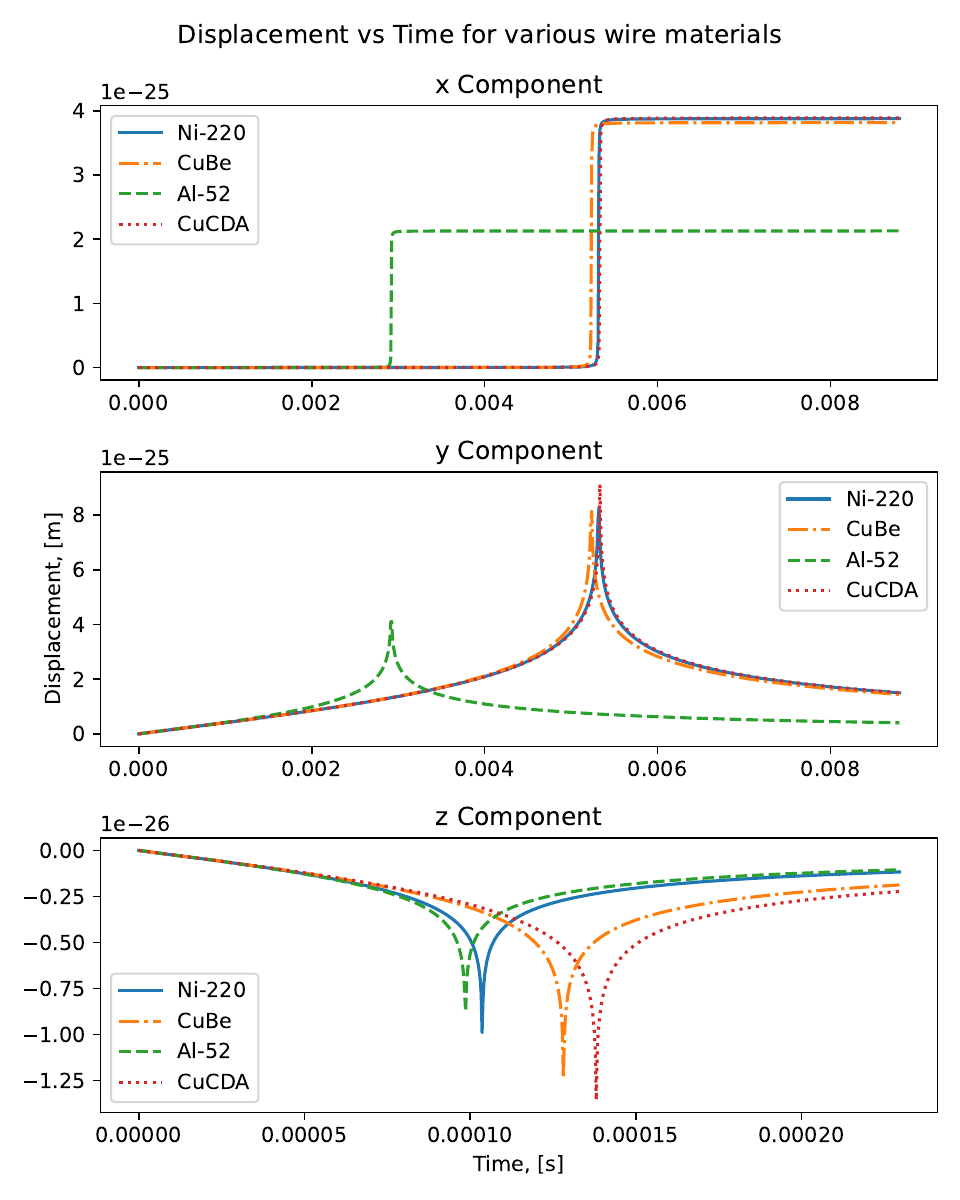}
    \caption{Time evolution of a single point on a 2 meter wire with a radius of 45 microns made of Copper Berylium (CuBe), Aluminum 5052 (Al-52), Copper CDA (CuCDA), and Nickel 220 (Ni-220). Displacement caused by a galactic orbit Planck scale dark matter particle traveling the same track as Fig.\ \ref{fig:impulseEx1}}
    \label{fig:diffWiresTime}
\end{figure}

As can be seen in Eqs.\ \eqref{eqn:dispLeft} and \eqref{eqn:dispRight}, the amplitude of the displacement is inversely proportional to wave speed, $w$. 
From Eqs.\ \eqref{eqn:dispLeft} and \eqref{eqn:dispRight} it follows that in order to achieve lower wave speeds we want a thin wire, with low Young's modulus, and large mass density, held at relatively low tension. 
In Fig.\ \ref{fig:diffWiresTime} we compare displacements expected from some commercially available wire options offered by California Fine Wire.\cite{CalFineWire} 
The simulated wires were 2 meters long, with a radius of 45 microns, and held at the same tension. 
Overall, the impact of wire choice on magnitude of displacement is fairly small, within the same order of magnitude, so any choice of thin and light wire should produce similar results.\cite{CalFineWire} 

For the calculations in this paper the Copper CDA wire was chosen, with a constant tension equivalent to ten times the wire's total weight to limit sagging (for a horizontal wire) or variation of actual tension along the wire (for a vertical wire).
This gives us a transverse wave speed of $14$ m/s and a longitudinal wavespeed of $3.6$ km/s.
Interestingly, choosing tension by this method eliminates the dependence of transverse wave speed on linear mass density seen in Eq.\ \eqref{eqn:transWaveSpeed}. The transverse wave speed is now primarily determined by the length of the wire, $l$,
\begin{align}
    \nonumber T &= 10 \cdot \mu gl\\
    w_s &= \sqrt{\frac{T}{\mu}} = \sqrt{10gl},
    \label{eqn:transverseMuRelation}
\end{align}
where $g$ is the Earth's gravitational acceleration on the wire.

This allows transverse displacement to be the same for a given length regardless of the wire chosen. 
The longitudinal wave speed (Eq.\ \ref{eqn:longWaveSpeed}) remains independent of the wire's length.

The wire is held under constant tension with fixed ends. Using this framework for our analysis we can find the displacement caused by various particles with different parameters.

Given the resolutions of commercially available, nanometer scale capacitive and piezo-resistance strain gauge displacement sensors, our goal is displacements of at least $2.4$ nm for transverse waves and $0.49$ nm for longitudinal waves.\cite{FLEMING2013106, ZHOU2025115988}

Micro-opto-electro-mechanical systems (MOEMS) have been noted to reach resolutions on the scale of picometers/Hz$^{1/2}$ and even femtometers/Hz$^{1/2}$.\cite{mi15081011, Anetsberger2009}
This means that we can roughly estimate the smallest detectable displacement as somewhere around $10 \times 10^{-15}$ m given the ms and sub-ms time scale of the wave pulses.
Displacements this small approach the standard quantum limit, however it has been shown that sub-SQL sensitivity is possible.\cite{Mason2019, Anetsberger2009} 

\subsection{Galactic Orbit Dark Matter}\label{sec:GODM}
Given the galactic orbit speed we can calculate the necessary mass to create nanometer scale displacements, shown in Table \ref{tab:galOrbitDM} where $\psi_x$ and $\psi_z$ are calculated with $\theta=\pi/2$ and $\psi_y$ is calculated with $\theta=\pi/4$ for the maximum possible displacement. 
The impact parameter $b$ is varied from 0.1 to 100 mm.

\begin{table*}[ht]
    \centering
    \caption{Minimum particle mass required for nanometer scale displacement of 2.4 nm in x and y and 0.49 nm in z. 
    $\theta=\pi/2$ for x and z. 
    $\theta=\pi/4$ for y.}
    \begin{tabular*}{\textwidth}{@{\extracolsep{\fill}} c c c c }
        \hline
        \hline
         b [mm] & \multicolumn{3}{l}{Minimum mass to incite detectable displacements}\\
         & in $x$  $\left[\times 10^{7}\text{ kg}\right]$&  in $y$ $\left[\times 10^{7}\text{ kg}\right]$ & in $z$ $\left[\times 10^{8} \text{ kg}\right]$\\   
        
         0.1& 3.69& 1.37& 6.93\\
         
         1&  3.69&  1.88& 9.39\\
        
         10&  3.73&  3.00& 14.52 \\
        
         100&  4.18&  7.33& 31.83\\
         \hline
         \hline
    \end{tabular*}
    \label{tab:galOrbitDM}
\end{table*}

From Table \ref{tab:galOrbitDM}, we would need a galactic orbit dark matter particle with mass on the order of $10^7$ kg or $10^{33}$ GeV/$c^2$ for transverse waves to be detectable, and at least another order of magnitude greater to produce detectable longitudinal waves. 

It is important to note that such mass is not plausible for an 'elementary' particle because it exceeds the limit on Planck scale dark matter by at least 15 orders of magnitude.\cite{Billard_2022, Adhikari2022, Aalbers2024, Aprile2023}
Planck scale dark matter, moving at galactic orbit speed, would have a flux of 
\begin{equation}
   \Phi_{DM} = \frac{v \rho_{DM}}{E_{DM}} = \frac{v}{M}\frac{\rho_{DM}}{c^2} \simeq 0.4 / \text{yr} / \text{m$^2$},
\end{equation}\label{eqn:particleFlux}
which is about $1.5$ million particles passing through Earth every second and would result in displacements on the scale of $10^{-24} \text{ to } 10^{-26}$ meters.

Applying Eq.\ \ref{eqn:particleFlux} to the values in Table \ref{tab:galOrbitDM} gives us a particle flux of $\sim 2 \times 10^{-16} / \text{yr}/\text{m}^2$, or 1 earthly event every 45 years.
A frequency of 1 earthly event per year (corresponding to a DM mass of $\sim10^{32}$ GeV/$c^2$ or $\sim 23$ kg) would produce displacements on the order of $10^{-10}$ to $10^{-12}$ m, however such a particle is still significantly greater than our upper limit on 'elementary' dark matter.

Given our current understanding of dark matter and the limits of displacement sensing, detecting dark matter using waves propagated along a wire as a medium is impossible.

\subsection{Neutrons}
Given the rarity of plausible dark matter events it would be ideal if we could test the detector using neutrons. They would have no electromagnetic interaction with our wire and there are various methods of generating neutrons and moderating their speeds. However, given a neutron's mass of $1.67 \times 10^{-27}$ kg we find that there is no plausible speed they could travel to create meaningful displacement in the wire. Even in a cryogenic environment, such as 0.1 K, where a neutron will be moving at only 35 m/s we would not see displacements with an order of magnitude greater than $10^{-40}$ m.
Note that this particle speed is significantly slower than our longitudinal wavespeed of $3.6$ km/s and thus we would not be able to make the assumptions in Sec.\  \ref{sec:impulse} that allow us to approximate the interaction as an impulse for the longitudinal wave component.

\subsection{Charged Particles} 
Finally, we examine if the design may be applicable to charged particle detection.
A massive particle acting on a wire of uniform mass is similar to a charged particle interacting with an oppositely charged static wire. 
Both forces can be described using the inverse square law, differing only by the leading factors. 
For a charged particle this would be 
\begin{equation}
    \frac{Q \delta q}{4 \pi \epsilon_0}.
\end{equation}
Here $Q$ is the particle's charge, $\delta q$ is the charge held by a small segment of the wire, and $\epsilon_0$ is the vacuum permittivity.

Following the same logic as Sec.\ \ref{sec:waveSolutions}, we can see that, for a charged particle, dividing impulse by twice the wire's differential mass, $\delta m$, results in a leading factor of
\begin{equation}
    \frac{Q}{8 \pi \epsilon_0} \frac{\delta q}{\delta m}.
\end{equation}
Additionally,
\begin{equation}
  \frac{\delta q}{\delta m} = \frac{\lambda}{\mu}  
\end{equation}
where $\lambda$ is linear charge density and $\mu$ is linear mass density. 
This means that for charged particles the ratio of charge density over mass density will affect our expected wave solutions. 

Additionally, unlike mass, charge can be either positive or negative and thus both attractive and repulsive interactions are possible in the case of charged particles interacting with a charged wire.
Finally, for non-static situations such as with a conducting wire, the charge density will change, distributing so that charge of the same sign as the interacting particle gathers on the ends of a wire.
Given a wire and particle with oppositely signed charge, we would expect it to enhance the displacement, however, because of the non-uniform distribution, these displacements would not follow the equations used in this paper.

Examining the interaction under an electrostatic situation, the most straightforward and similar example is a wire with uniform charge density.
If we wrap the wire in a cylindrical shell and induce a voltage drop of $V$ across the two, we can create a cylindrical capacitor.
Given $r$ as the wire's radius and $R$ as the radius of the cylindrical shell we can find linear charge density, in the approximation of the capacitor being infinite, as
\begin{equation}
    \lambda = \frac{2\pi \epsilon_0 V}{\ln{\left(R/r\right)}}.
\end{equation}
Note that $r$ is related to linear mass density by $\mu = \rho\pi r^2$ and thus 
\begin{equation}
    \frac{Q}{8 \pi \epsilon_0} \frac{\delta q}{\delta m} = 
    \frac{Q}{4 \pi \rho} \frac{V}{r^2\ln{\left(R/r\right)}}.
    \label{eqn:coulombLeadingFactor}
\end{equation}

Replacing $GM$ in Eqs.\ \ref{eqn:dispLeft} and \ref{eqn:dispRight} with the leading factor defined in Eq.\ \ref{eqn:coulombLeadingFactor}, we can calculate the displacements a particle of charge $Q$ would cause.
Using a wire with radius $r = 45$ microns, a voltage of 3 kV, and a shell of radius $R = 2$ cm, a particle passing within $1$ mm of the wire, within the shell's radius, with a charge of $-e$, would cause transverse displacements on the order of $10^{-22}$ to $10^{-24}$ m at light-speed. 

It should be noted that the interaction will be limited to the time the particle is within the bounds of capacitor, and thus the wave shape may be different. However, as the interaction is strongest at very close distances and the shell of the capacitor is over ten times larger than the impact parameter of the particle used, it should still serve as a useful estimate of order of magnitude.

Slower charged particles would produce larger displacements. Similar to the neutron case described above, there is a range of particle speeds for which the assumptions in Sec.\ \ref{sec:impulse} are violated for the longitudinal wave component but are still valid for the transverse component. We find that a charged particle moving at 100 m/s would produce a transverse displacement on the order of femtometers, comparable to the resolution of the most sensitive displacement sensors currently available. \cite{mi15081011, Anetsberger2009}

% \section{Author Declarations}
% The authors have no conflicts to disclose

\bibliography{aip}

\end{document}